\documentclass{article}\usepackage[]{graphicx}\usepackage[]{color}
\makeatletter
\def\maxwidth{ %
  \ifdim\Gin@nat@width>\linewidth
    \linewidth
  \else
    \Gin@nat@width
  \fi
}
\makeatother

\usepackage{Sweave}

\usepackage[hyphens]{url}
\usepackage[utf8]{inputenc}
\usepackage[sort&compress, numbers]{natbib}
\usepackage{graphicx}
\usepackage{authblk}
\usepackage[english]{babel}
\usepackage{setspace}
\usepackage{fancyhdr}
\usepackage{array}
\usepackage[margin=1in]{geometry}
\usepackage{float}
\usepackage{datetime}
\usepackage{amsmath}
\usepackage{amssymb}
\usepackage{bbm}
\usepackage{bm}
\usepackage[final]{pdfpages}
\usepackage{longtable}
\usepackage{chngcntr}
\usepackage{placeins}
\usepackage{microtype}
\usepackage{tikz}
\usepackage{makecell}
\usepackage{hyperref}
\usepackage{subcaption}
\usepackage{rotating}
\usepackage{soul}
\usepackage{tabularx}
\usepackage[T1]{fontenc}
\usepackage{lmodern}

\hypersetup{
    colorlinks = true,
    citecolor = {blue},
    urlcolor = {blue},
    menucolor = {blue},
    linkcolor = {blue}
}

\newcommand\code{\text}
\def\@codex#1{{\normalfont\ttfamily\hyphenchar\font=-1 #1}\egroup}
\let\proglang=\textsf
\newcommand\pkg{\textbf}

\makeatletter
\def\@author{}
\renewcommand\@author{\ifx\AB@affillist\AB@empty\AB@author\else
      \ifnum\value{affil}>\value{Maxaffil}\def\rlap##1{##1}%
    \AB@authlist \\*[0.1cm] \small on behalf of EU-PEARL (EU Patient-cEntric clinicAl tRial pLatforms) Consortium \\[\affilsep]\AB@affillist
    \else  \AB@authors\fi\fi}
\makeatother

\title{CohortPlat: Simulation of cohort platform trials investigating combination therapies}

\author[1]{Elias Laurin Meyer}
\author[2]{Peter Mesenbrink}
\author[3]{Cornelia Dunger-Baldauf}
\author[3,4]{Ekkehard Glimm}
\author[1,*]{Franz König}
\affil[1]{Center for Medical Statistics, Informatics, and Intelligent Systems, Medical University of Vienna, Austria}
\affil[2]{Novartis Pharmaceuticals Corporation, One Health Plaza, East Hanover, NJ, USA}
\affil[3]{Novartis Pharma AG, Basel, Switzerland}
\affil[4]{Institute of Biometry and Medical Informatics, University of Magdeburg, Germany}
\affil[*]{Correspondence: franz.koenig@meduniwien.ac.at; Tel.: +43-1-40400-74800}
\date{}         
\IfFileExists{upquote.sty}{\usepackage{upquote}}{}
\begin{document}

\maketitle

\begin{abstract}

Platform trials have gained a lot of attention recently as a possible remedy for time-consuming classical two-arm randomized controlled trials, especially in early phase drug development. This short article illustrates how to use the \pkg{CohortPlat} \proglang{R} package to simulate a cohort platform trial, where each cohort consists of a combination treatment and the respective monotherapies and standard-of-care. The endpoint is always assumed to be binary. The package offers extensive flexibility with respect to both platform trial trajectories, as well as treatment effect scenarios and decision rules. As a special feature, the package provides a designated function for running multiple such simulations efficiently in parallel and saving the results in a concise manner. Many illustrations of code usage are provided.

\end{abstract}


\section{Introduction} \label{sec:intro}

In recent years, master protocol and especially platform trials have gained a lot of momentum and are considered by many as the future of clinical drug development programs (\citet{Woodcock2017}, \citet{Park2019}, \citet{ballarini2021optimizing}). Platform trials facilitate the investigation of multiple treatments in multiple subgroups and can be seen as an extension to adaptive multi-arm multi-stage (MAMS) trials, where treatments and sub-studies can enter and leave over time according to pre-defined advancement and exit rules (\citet{Angus2019}). Compared to more traditional trial designs and MAMS trials, platform trials add many additional layers of flexibility and therefore complexity, such as staggered entry of treatments and substudies over time, more flexible data sharing options across treatments arms and changing allocation ratios (\citet{Park2020}), which makes simulation of such trials considerably more difficult. At the same time, simulation of such complex trials incorporating all possible sources of flexibility is crucial in understanding how such trials perform in real life. Many statistical challenges with respect to platform trials remain, e.g. when and how to correct for multiple testing (\citet{Wason2014_2}, \citet{Howard2018}, \citet{Stallard2019}, \citet{Collignon2020}, \citet{posch2020commentare}, \citet{bretz2020commentary}, \citet{burger2021use}), which is one of the topics we are working on in within the EU-PEARL project \url{https://www.imi.europa.eu/projects-results/project-factsheets/eu-pearl}. As the focus of this article is on the task of simulating platform trials, we refer the readers to \citet{Meyer2020} for a comprehensive overview of current methodology and statistical issues related to platform trials.

A systematic review identifying and comparing available software for the design and simulation of MAMS and platform trials has been published recently (\citet{Meyer2021}). Some software exists which can potentially be adapted to be used in the simulation of platform trials, such as the \proglang{R} \citep{R} package \pkg{MAMS} (\citet{jaki2019r}), however there exists no software to date which incorporates most or all of the core features of platform trials. Previous simulation studies have mostly focused on particular aspects of the platform trial, such as \citet{Saville2016}, \citet{Yuan2016}, \citet{Ventz2017}, \citet{Hobbs2018_2}, \citet{Tang2018} and \citet{Wason2020}, however no (available) software to date has tied all the features together. A few online applications exist for simulating MAMS trials or simple platform trials, described in more detail in \citet{Grayling2020}, \citet{Thorlund2019} and at \url{https://rpact.shinyapps.io/public/}. Another \proglang{R} package called \pkg{Octopus} exists on GitHub (\cite{Wathen2020}), which is a modular platform trial simulator, where users can re-write parts of the source code to achieve flexibility with respect to some of the platform trial features. However, we have found it to be of limited use out of the box, due to the limited documentation and prerequisites to re-write source code.

In this paper we describe the use of the \proglang{R} package \pkg{CohortPlat}, which was developed to facilitate simulation of a particular type of platform trial design. The trial design in question was described in more detail by the authors in a recently submitted paper, which is available on arXiv as a preprint (\citet{Meyer2021decision}). The \pkg{CohortPlat} package is available from CRAN at https://CRAN.R-project.org/package=CohortPlat.

\section{Trial Design}

\begin{figure}[ht]
\centering
\begin{subfigure}{1\textwidth}
\centering
\includegraphics[width=0.777\textwidth]{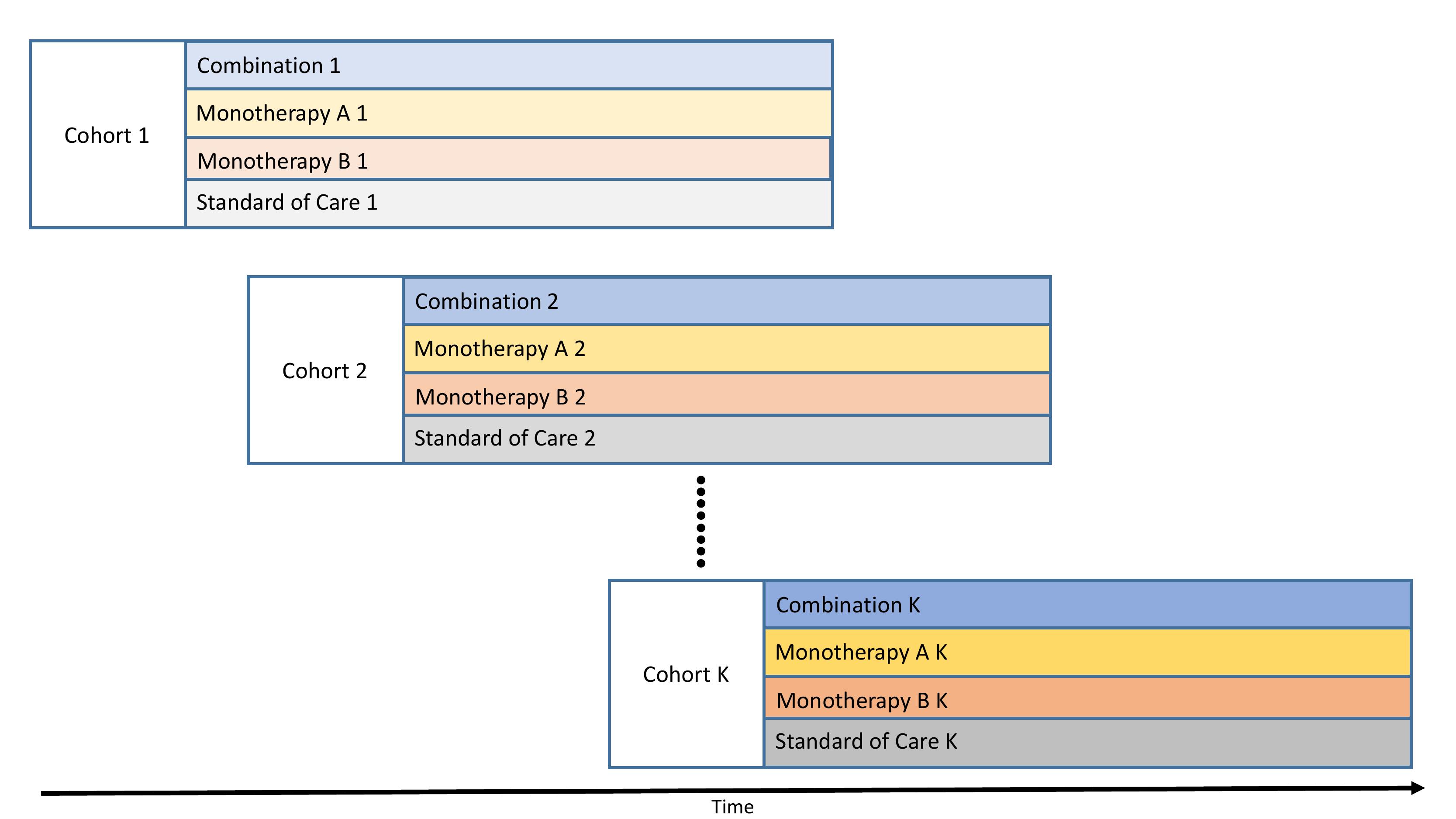}
\caption{Separate Cohorts} \label{fig:trialdesign1}
\end{subfigure}

\bigskip
\begin{subfigure}{1\textwidth}
\centering
\includegraphics[width=0.777\textwidth]{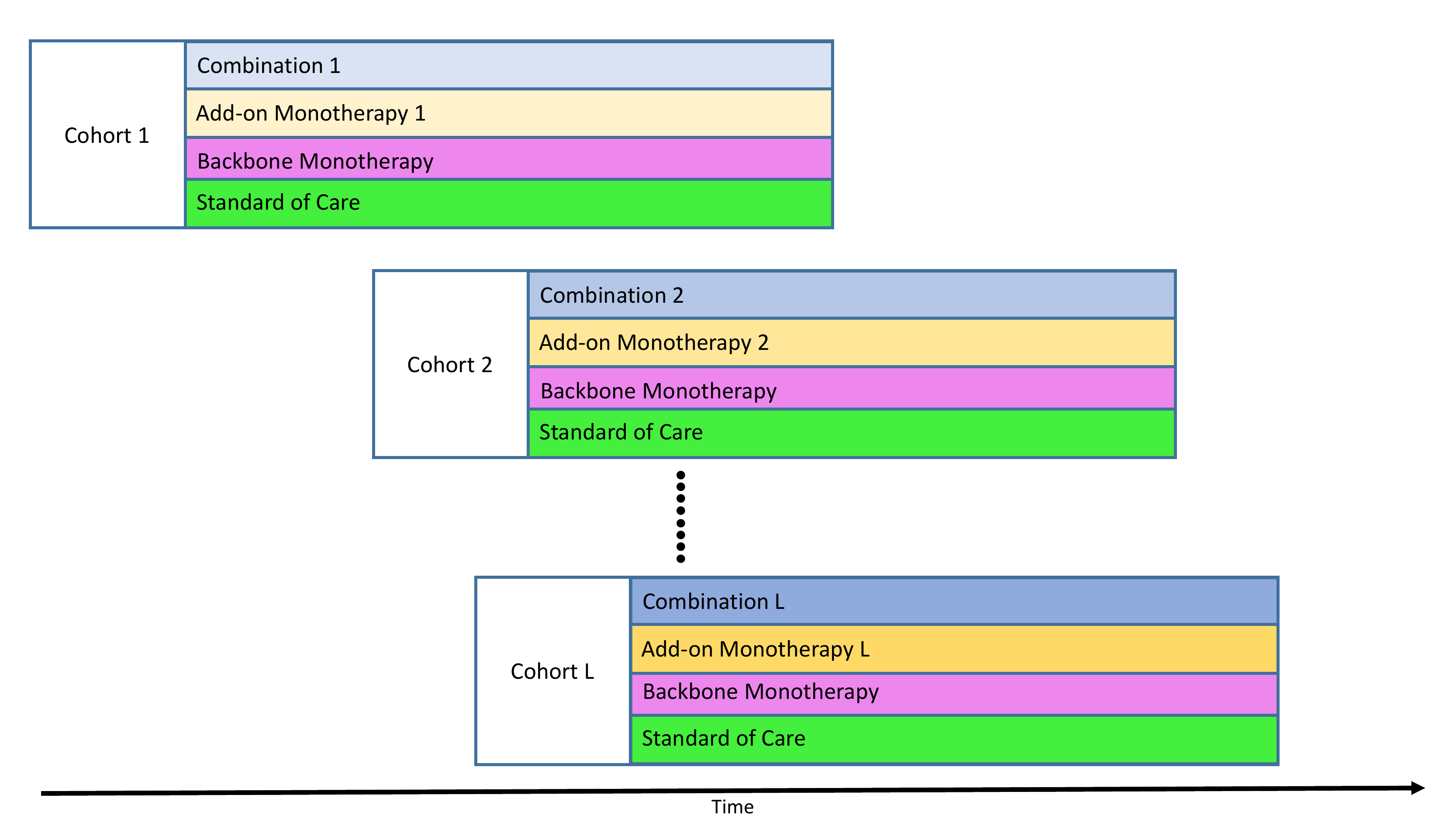}
\caption{Shared Arms} \label{fig:trialdesign2}
\end{subfigure}
\caption{Schematic overview of the proposed platform trial design. New cohorts consisting of a combination therapy arm, the two respective monotherapy arms and a standard-of-care (SoC) arm randomly enter the platform over time. In the simple case depicted in figure \ref{fig:trialdesign1}, the cohorts consist of separate SoC arms and separate monotherapies, and as a result separate combination therapies (as indicated by the differently shaded colors). In figure \ref{fig:trialdesign2}, we consider the special case where one of the monotherapy arms is used in every cohort (referred to as "backbone monotherapy") and also the SoC is the same in every cohort (as indicated by the same colors). The second monotherapy (referred to as "add-on monotherapy") and therefore the combination therapy are different in every cohort (as indicated by the differently shaded colors). In this setting, sharing of SoC and backbone monotherapy data across cohorts might be desired. While in both figures colors are used to indicate that treatments are either the same (if the color is the same) or belong to the same class of treatment (different shades), patients are different for every treatment and every cohort. As an example, in figure \ref{fig:trialdesign2} the 8 different depicted treatments are investigated in 12 sets of patients (number of bars).}
\label{fig:trialdesign}
\end{figure}

We look at an open-entry, cohort platform study design with a binary endpoint investigating the efficacy of a two-compound combination therapy compared to the respective monotherapies and standard-of-Care (SoC). After an initial inclusion of one or more cohorts, we allow new cohorts to enter the trial over time. Each cohort consists of up to four arms: combination therapy, monotherapy A, monotherapy B and an optional SoC, of which monotherapy B and SoC could be the same in all cohorts (see section \ref{sec:assumptions_parameters}). We furthermore assume to conduct one interim analysis for efficacy and futility for every cohort. A schematic overview of the trial design can be found in Figure \ref{fig:trialdesign}, more information is provided in \citet{Meyer2021decision}.

The package was developed considering a clinical drug development program including the phase III of drug approval. For this purpose, the package includes decision rules for declaring superiority of the combination therapy over both monotherapies, and monotherapies over SoC. The package facilitates investigation of the operating characteristics (type I error and power) of these decision rules. Depending on the level of prior study information available, i.e. whether or not the superiority of the monotherapies over SoC has already been shown, we differentiated three testing strategies. In the first testing strategy, we assume that the monotherapies are already approved, therefore we are only interested in testing the combination therapy against both monotherapies. In the second testing strategy, we assume superiority of one of the monotherapies versus SoC has been shown, but not for the second monotherapy. Therefore, compared to the first testing strategy, we additionally test the unestablished monotherapy against SoC, resulting in three comparisons. In the third and last testing strategy, we do not assume any superiority has been shown yet. Therefore, compared to the second testing strategy, we test both monotherapies against SoC, resulting in four comparisons.

For every testing strategy a certain number of comparisons are performed. For every comparison, we allow multiple simultaneous decision rules, which can be either Bayesian based on the posterior distributions of the response rates of the respective study arms, or frequentist based on either a p-value or point estimates and confidence intervals. Generally, we allow decision rules for declaring efficacy and for declaring futility.

\section{Simulation assumptions and parameters} \label{sec:assumptions_parameters}

\subsection{Cohort Structure}

In general, all cohorts consist of two monotherapy arms and a combination arm of the two monotherapies. As described in figure \ref{fig:trialdesign}, a special case that can be simulated using the package is when the SoC arm and one of the monotherapies is the same in every cohort (referred to as "backbone monotherapy"). This could be the case when e.g. all cohorts target the same patient populations and one monotherapy has putative or already established efficacy over SoC. In this case, also sharing of data across cohorts might be desired. There are four options for the SoC arm: 1) All cohorts always include a SoC arm. 2) All cohorts start with a SoC arm, but after the first time a monotherapy has been found to be superior to SoC, newly enrolled cohorts do not have a SoC arm. 3) Same as 2), but only for one of the backbone monotherapy. 4) No cohort has a SoC arm. Depending on the cohort structure, the testing strategy changes, e.g. when no SoC arm is included in a cohort, only two comparisons - combination therapy versus monotherapies - are conducted.

\subsection{Treatment efficacy assumptions}

In addition to specifying the trial design parameters, treatment efficacy assumptions need to be made. For every new cohort that enters the platform trial, we allow the response rates of the respective arms to be drawn independently from previously specified discrete probability distributions, which should capture the observed heterogeneous treatment effects. Several options are possible: Either response rates are drawn directly, or risk-ratios, risk-differences or odds-ratios are drawn based on SoC as reference arm. Furthermore, odds-ratios can be drawn randomly for the two monotherapies and subsequently an interaction effect is randomly drawn for the combination therapy. Some examples are shown in section \ref{sec:illustrations} and an in-detail description is provided in the appendix.

\subsection{Endpoints and sample sizes}

For every cohort, after a certain number of patients have been enrolled, we conduct an interim analysis for early efficacy and futility. If the cohort is not stopped at interim, we further recruit until a pre-defined number of patients have been enrolled. At interim, we either use the primary endpoint, or a surrogate endpoint with a certain correlation (e.g. if the primary endpoint is a long-term endpoint and a short-term surrogate is available). When simulating the binary interim and final endpoints for each patient, we randomly draw from a multinomial distribution with four possible outcome pairs: 0/0, 0/1, 1/0 and 1/1, where the first number corresponds to the interim outcome and the second number corresponds to the final outcome. The probabilities with which these outcomes are drawn are anchored in the true response rate for the final endpoint (which we will denote by $rr$, but which depends on the cohort and arm). In general, the package allows for arbitrary functions that take the true response rate for the final endpoint as input and return a vector of length 4 corresponding to the joint probability distribution function mentioned previously. By placing no probability mass on the 0/1 and 1/0 options, we could achieve a scenario in which we use the final outcome also at interim. An example is provided in section \ref{sec:illustrations}.

\subsection{Block Randomization}

A block randomization for all cohort active at any current point in time was implemented. To give an example: If at any point of the simulation there are two currently enrolling cohorts with the allocation ratios 2:2:1:1 (combination therapy : monotherapy A : monotherapy B : SoC), then in every iteration 6 patients are allocated to each of the cohorts (i.e. in this example 12 patients were simulated in one iteration). Within each of the cohorts, these 6 patients are placed on the treatment arms according to the allocation ratio, i.e. 2 patients are allocated to the combination therapy, 2 patients are allocated to monotherapy A, 1 patient is allocated to the monotherapy B and 1 patient is allocated to SoC. The recruitment speed is therefore directly influenced by the number of and allocation ratio within the currently ongoing cohorts. As another consequence, the planned interim and final sample sizes can be slightly overshot.

\subsection{Target Product Profile}

When calculating the operating characteristics of a trial design and the chosen decision rules, we need a notion of whether or not a correct or wrong decision was made. We want to assess this via the target product profile, which specifies how much better (in terms of risk-difference, risk-ratio or odds-ratio) the combination therapy needs to perform than both of the monotherapies and how much better the monotherapies need to perform compared to the SoC for a positive decision to be considered a true positive decision. Consequently, if any of these conditions is not met, a positive decision would be considered a false positive decision. Analogously, true and false negative decisions are defined. For more details see \citet{Meyer2021decision}.

\subsection{Level of Data Sharing Across Cohorts}

In the alternative study design discussed earlier, where SoC and one of the monotherapies are the same in every cohort, and related to the flexible decision rules described later, we can specify whether we want to share information on the backbone monotherapy and SoC arms across the study cohorts. We allow four options: 1) no data sharing, using only data from the current cohort, 2) sharing concurrently collected data, 3) using a dynamic borrowing approach further described in \citet{Meyer2021decision} in which the degree of shared data increases with the homogeneity of the treatment efficacy, i.e. sharing more, if the treatment efficacy is similar to previous cohorts and 4) full sharing (i.e. pooling) of all available data.

\subsection{Further Platform Trial Rules}

Cohorts enter the platform trial over time at random time points. This is controlled via a probability to include a new cohort after every included patient. Similarly, a probability to stop a cohort for safety reasons after every included patient can be specified. Furthermore, the maximum number of cohorts allowed to enter the platform can be limited.

While in theory a platform trial could run perpetually, it makes sense to foresee stopping it at some point, e.g. after a certain number of successful cohorts or at a maximum sample size. Maybe the trial sponsors want to stop it immediately as soon as they have identified a successful combination therapy. The number of successful cohorts after which the platform stops instantly can be set to particular number or to Infinity, meaning all cohorts that enter the platform will always finish evaluating. Another option is to set a maximum sample size, after which recruitment to all available cohorts will be stopped. Lastly, it can be controlled whether or not to allow new cohorts to enter the platform trial once a cohort has been declared successful.

\clearpage

\section{Decision Rules} \label{sec:decisionrules}

In general, the user is required to specify the exact decision criteria that should be used for superiority and futility for all four possible comparisons: combination therapy vs. monotherapy A, combination therapy vs. monotherapy B, monotherapy A vs. SoC and monotherapy B vs. SoC, further differentiating whether the conducted analysis is an interim analysis or the final analysis. Several different types of decisions rules are supported: Bayesian decision rules based on posterior probabilities for two arms for superiority (\code{Bayes\textunderscore Sup}) and futility (\code{Bayes\textunderscore Fut}) or single arm for superiority (\code{Bayes\textunderscore SA\textunderscore Sup}) and futility (\code{Bayes\textunderscore SA\textunderscore Fut}) and frequentist decision rules based on a frequentist test for superiority (\code{P\textunderscore Sup}) and futility (\code{P\textunderscore Fut}), as well as decision rules on parameter estimates (\code{Est\textunderscore Sup\textunderscore Fut}) and their confidence intervals (\code{CI\textunderscore Sup\textunderscore Fut}). The decision rules are passed as a list consisting of two sublists (one for the interim and one for the final decision making). The requirement for all lists of decision rules are the same: the list needs to consist of one of the respective decision rule's elements (described below), i.e. matrices for \code{Bayes\textunderscore Sup}, \code{Bayes\textunderscore Fut}, \code{Bayes\textunderscore SA\textunderscore Sup} , \code{Bayes\textunderscore SA\textunderscore Fut} and lists for \code{P\textunderscore Sup} \code{P\textunderscore Fut}, \code{Est\textunderscore Sup\textunderscore Fut} and/or \code{CI\textunderscore Sup\textunderscore Fut}. For each of the decision rules we need to follow a certain structure which is explained in the following sub-sections. Decision rules can be arbitrarily combined (e.g. Bayesian superiority and frequentist futility rules), do not have default values and are not checked for consistency. In order to declare efficacy, all specified efficacy decision rules must be simultaneously fulfilled. In order to declare futility, it is enough if any of the specified futility decision rules is fulfilled. It is possible that neither is satisfied. If this happens at the final analysis, we declare the combination therapy unsuccessful, but due to not reaching the superiority criterion at the maximum sample size, and not due to reaching the futility criterion. While this is only a technical difference, this information should be available when conducting the simulations. At interim, early futility or efficacy stopping is possible.

\subsection{Bayesian Superiority Rules} \label{sec:decisionrules_bayessup}

For \code{Bayes\textunderscore Sup}, the number of rows of the matrix determines how many criteria need to be simultaneously true to declare superiority. The first column refers to the required superiority margin and the second column to the required confidence. The third column, which is used for the concept of a "promising" drug gives the threshold at which we declare a drug "promising", in case we did not declare it superior. This concept of a "promising" treatment is currently only used in the simulation output for manual investigation, but it was foreseen that users might want to add e.g. different interim decisions for "promising" treatments in the simulation code. To set a superiority decision rule of the form: GO, if $P(X>Y+0.1)>0.8$ and declare promising, if $P(X>Y+0.1)>0.5$, use a matrix with one row and three columns (first column corresponds to margin $0.1$, second column to superiority threshold $0.8$ and third column to "promising" margin $0.5$):

\begin{Schunk}
\begin{Sinput}
R> B1 <- matrix(nrow = 1, ncol = 3)
R> B1[1,] <- c(0.10, 0.80, 0.5)
\end{Sinput}
\end{Schunk}

The margin can be set to a value smaller than 0, thereby implementing a Bayesian non-inferiority decision rule.

\subsection{Bayesian Futility Rules} \label{sec:decisionrules_bayesfut}

For \code{Bayes\textunderscore Fut}, the number of rows of the matrix determines how many criteria need to be simultaneously true to declare futility. The first column refers to the required superiority margin and the second column to the required confidence. To set a futility decision rule of the form: STOP, if $P(X>Y+0.1)<0.5$, use a matrix with one row and two columns:

\begin{Schunk}
\begin{Sinput}
R> B2 <- matrix(nrow = 1, ncol = 2)
R> B2[1,] <- c(0.10, 0.50)
\end{Sinput}
\end{Schunk}

\subsection{Bayesian Single Arm Superiority Rules} \label{sec:decisionrules_bayessasup}

For \code{Bayes\textunderscore SA\textunderscore Sup}, the number of rows of the matrix determines how many criteria need to be simultaneously true to declare superiority. The first column refers to the required value and the second column to the required confidence. The third column, which is used for the concept of a "promising" drug gives the threshold at which we declare a drug "promising", in case we did not declare it superior. Imagine we have the following two superiority criteria: GO, if ($P(X>0.05)>0.7$ AND $P(X>0.1)>0.5$). In this case, we do not want to use the concept of "promising", which we can achieve by choosing a boundary that will never be crossed. To set these decision rules, we use a matrix with two rows and three columns:

\begin{Schunk}
\begin{Sinput}
R> B3 <- matrix(nrow = 2, ncol = 3)
R> B3[1,] <- c(0.05, 0.70, 1.00)
R> B3[2,] <- c(0.10, 0.50, 1.00)
\end{Sinput}
\end{Schunk}

\subsection{Bayesian Single Arm Futility Rules} \label{sec:decisionrules_bayessafut}

For \code{Bayes\textunderscore SA\textunderscore Fut}, the number of rows of the matrix determines how many criteria need to be simultaneously true to declare futility. The first column refers to the required value and the second column to the required confidence. To set a futility decision rule of the form $P(X>0.1)<0.5$, use a matrix with one row and two columns:

\begin{Schunk}
\begin{Sinput}
R> B4 <- matrix(nrow = 1, ncol = 2)
R> B4[1,] <- c(0.10, 0.50)
\end{Sinput}
\end{Schunk}

\subsection{Frequentist Superiority Rules} \label{sec:decisionrules_psup}

For \code{P\textunderscore Sup}, the number of list elements determines how many criteria need to be simultaneously true to declare superiority. Each list element needs to follow the following structure: The first element passes the testfunction. The testfunction may be any R function taking a 2*2 table as input and returning a object with an element called "p.value". The second element passes the significance level to declare superiority, the third element passes the "promising" significance level which is used for the concept of a "promising" drug. The fourth element decides whether a Bonferroni correction (assuming two tests, therefore taking half of the specified significance levels) should be used. As an example, consider the use of a one-sided Chi-squared test for the GO decision with a level of 2.5\% without continuity correction using Bonferroni multiplicity correction, and not using the "promising" concept (i.e. choosing a boundary that will never be crossed). In this case, the list would look like this:

\begin{Schunk}
\begin{Sinput}
R> P1 <- list(list(
+    testfun = function(x) stats::prop.test(x, alternative = "less",
+                                           correct = FALSE),
+    p_sup = 0.025, p_prom = 0, p_adj = "B"))
\end{Sinput}
\end{Schunk}

\subsection{Frequentist Futility Rules} \label{sec:decisionrules_pfut}

For \code{P\textunderscore Fut}, the number of list elements determines how many criteria need to be simultaneously true to declare futility. Each list element needs to follow the following structure: The first element passes the testfunction with the same properties as described above for \code{P\textunderscore SUP}. The second element passes the significance level to declare futility. The third element decides whether a Bonferroni correction (assuming two tests, therefore taking half of the specified significance levels) should be used. To set a futility decision rule of the form: STOP, if $p \geq 0.5$, with p stemming from a one-sided Chi-Square test without continuity correction using no multiplicity correction, use a list:

\begin{Schunk}
\begin{Sinput}
R> P2 <- list(list(
+    testfun = function(x) stats::prop.test(x, alternative = "less",
+                                           correct = FALSE),
+    p_fut = 0.5, p_adj = "none"))
\end{Sinput}
\end{Schunk}

\subsection{Superiority and Futility Rules based on Parameter Estimates} \label{sec:decisionrules_estsupfut}

For \code{Est\textunderscore Sup\textunderscore Fut}, the number of list elements determines how many criteria need to be simultaneously true to declare superiority/futility. Each list element needs to follow the following structure: The first element passes the type of the point estimate. Options are "RR" (risk ratio) and "OR" (odds ratio). The second element passes the threshold to declare superiority. The third element passes the threshold to declare futility. The fourth element passes the promising threshold which is used for the concept of a "promising" drug. To set a rule of the form: GO, if RR $\geq 1.2$ and STOP, if RR $\leq 1$, use a list:

\begin{Schunk}
\begin{Sinput}
R> P3 <- list(list(est = "RR", p_hat_sup = 1.2, p_hat_fut = 1,
+                  p_hat_prom = Inf))
\end{Sinput}
\end{Schunk}

\subsection{Superiority and Futility Rules based on Confidence Intervals} \label{sec:decisionrules_cisupfut}

For \code{CI\textunderscore Sup\textunderscore Fut}, the number of list elements determines how many criteria need to be simultaneously true to declare superiority/futility. Each list element needs to follow the following structure: The first element passes the type of the point estimate. Options are "AR" (risk difference), "RR" (risk ratio) and "OR" (odds ratio). The second element passes the desired coverage probability of the confidence interval. Confidence intervals are calculated using the \code{riskratio.small()} and \code{odds.ration()} functions from the \pkg{epitools} package and the \code{binom.test()} function from the \pkg{stats} package. The third element passes the threshold for the upper confidence bound to declare superiority. The fourth element passes the threshold for the lower confidence bound to declare futility. The fifth element passes the promising threshold for the upper confidence bound which is used for the concept of a "promising" drug. To set a rule of the form: GO, if upper bound of 95\% CI for RR $\geq 1.2$ and STOP, if lower bound of 95\% CI for RR $\leq 1$, use a list:

\begin{Schunk}
\begin{Sinput}
R> P4 <- list(list(est = "RR", ci = 0.95, p_hat_lower_sup = 1.2,
+                  p_hat_upper_fut = 1, p_hat_lower_prom = Inf))
\end{Sinput}
\end{Schunk}

\section{Illustrations} \label{sec:illustrations}

Due to the multitude of simulation parameters that control the behavior of the platform trial, the assumptions about the treatment effect and the decision rules, careful thought needs to be given to the choice of these parameters. In this section, we will illustrate how to use the package to simulate a cohort platform trial using Bayesian decision rules. We will then extend this example to show how the package can be used to investigate multiple simulation assumptions. An example of how to set frequentist decision rules can be found in the appendix \ref{freq_dec}.

\subsection{Setting Bayesian decision rules}

Assume we want to implement the following superiority decision rules at the final analysis:

\begin{align*}
\text{GO, if } & (P(\pi_{Comb} > \pi_{MonoA} + 0.10 | Data) > 0.8) \ \wedge \\
               & (P(\pi_{Comb} > \pi_{MonoB} + 0.10 | Data) > 0.8) \ \wedge \\
               & (P(\pi_{MonoA} > \pi_{SoC} + 0.05 | Data) > 0.8) \ \wedge \\
               & (P(\pi_{MonoB} > \pi_{SoC} + 0.05 | Data) > 0.8) \\
\end{align*}

At interim, we use the same superiority decision rules to stop for early efficacy. Furthermore, we allow the following decision rules for early futility:

\begin{align*}
\text{STOP, if } & (P(\pi_{Comb} > \pi_{MonoA} | Data) < 0.6) \ \vee \\
                 & (P(\pi_{Comb} > \pi_{MonoB} | Data) < 0.6) \ \vee \\
                 & (P(\pi_{MonoA} > \pi_{SoC} | Data) < 0.6) \ \vee \\
                 & (P(\pi_{MonoB} > \pi_{SoC} | Data) < 0.6) \\
\end{align*}

Furthermore, we are not using the "promising" concept described in section \ref{sec:decisionrules}. The above decision rules are achieved by specifying the decision rules in the following way:

\begin{Schunk}
\begin{Sinput}
R> # Bayesian Superiority Rules
R> 
R> # Comparison Combo vs Monotherapy A
R> Bayes_Sup1 <- matrix(nrow = 1, ncol = 3)
R> Bayes_Sup1[1,] <- c(0.10, 0.80, 1.00)
R> # Comparison Combo vs Monotherapy B
R> Bayes_Sup2 <- matrix(nrow = 1, ncol = 3)
R> Bayes_Sup2[1,] <- c(0.10, 0.80, 1.00)
R> # Comparison Monotherapy A vs SoC
R> Bayes_Sup3 <- matrix(nrow = 1, ncol = 3)
R> Bayes_Sup3[1,] <- c(0.05, 0.80, 1.00)
R> # Comparison Monotherapy B vs SoC
R> Bayes_Sup4 <- matrix(nrow = 1, ncol = 3)
R> Bayes_Sup4[1,] <- c(0.05, 0.80, 1.00)
R> 
R> # Wrapup in package format
R> # The two sublists of Bayes_Sup are identical,
R> # because the same superiority rules are used
R> # at interim and at final. They could also differ.
R> Bayes_Sup <-
+    list(
+      list(
+        Bayes_Sup1,
+        Bayes_Sup2,
+        Bayes_Sup3,
+        Bayes_Sup4
+        ),
+      list(
+        Bayes_Sup1,
+        Bayes_Sup2,
+        Bayes_Sup3,
+        Bayes_Sup4
+        )
+      )
R> 
R> 
R> # Bayesian Futility Rules
R> 
R> # Comparison Combo vs Monotherapy A
R> Bayes_Fut1 <- matrix(nrow = 1, ncol = 2)
R> Bayes_Fut1[1,] <- c(0.00, 0.60)
R> # Comparison Combo vs Monotherapy B
R> Bayes_Fut2 <- matrix(nrow = 1, ncol = 2)
R> Bayes_Fut2[1,] <- c(0.00, 0.60)
R> # Comparison Monotherapy A vs SoC
R> Bayes_Fut3 <- matrix(nrow = 1, ncol = 2)
R> Bayes_Fut3[1,] <- c(0.00, 0.60)
R> # Comparison Monotherapy B vs SoC
R> Bayes_Fut4 <- matrix(nrow = 1, ncol = 2)
R> Bayes_Fut4[1,] <- c(0.00, 0.60)
R> 
R> # Wrapup in package format
R> Bayes_Fut <-
+    list(
+      list(
+        Bayes_Fut1,
+        Bayes_Fut2,
+        Bayes_Fut3,
+        Bayes_Fut4),
+      list(
+        Bayes_Fut1,
+        Bayes_Fut2,
+        Bayes_Fut3,
+        Bayes_Fut4
+        )
+      )
\end{Sinput}
\end{Schunk}

The simulation function expects a list with two sublists consisting of four matrices in the same dimensions as input for both \code{Bayes\textunderscore Sup} and \code{Bayes\textunderscore Fut}. If some comparisons are not to be conducted, the matrix elements can be \code{NA}. The exact naming is irrelevant.

\subsection{Assumptions and Design Parameters}

Some simulation parameters have default values (e.g. even when deciding to stop the platform trial, we will not stop it immediately, but rather let all currently active cohorts finish as planned). These can be accessed by typing \code{?simulate\textunderscore trial}. Furthermore, we want to fix the following assumptions and design choices:
\begin{itemize}
\item A probability to stop for safety for every patient of 0.01\%.

\begin{Schunk}
\begin{Sinput}
R> safety_prob <- 0.0001
\end{Sinput}
\end{Schunk}

\item We want to run the simulations with independent cohorts as in figure \ref{fig:trialdesign1}. To achieve the design illustrated in figure \ref{fig:trialdesign2}, the \code{sharing\textunderscore type} parameter can be set to \code{"all"}, \code{"concurrent"} or \code{"dynamic"}, referring to the type of data sharing that will be used.

\begin{Schunk}
\begin{Sinput}
R> sharing_type <- "cohort"
\end{Sinput}
\end{Schunk}

\item In order to declare a positive a true positive, we want the underlying response rate of the combination therapy to be at least 10\% points better than both monotherapies, and we want the monotherapies to be at least 5\% points better than SoC.

\begin{Schunk}
\begin{Sinput}
R> target_rr <- c(0.10, 0.05, 1)
R> # First element corresponds to superiority margin for combination
R> # Second element corresponds to superiority margin for monotherapies
R> # Third element corresponds to the scale chosen for the required
R> # superiority, e.g. "1" refers to risk-difference,
R> # "2" would refer to risk-ratio and "3" to odds-ratio
\end{Sinput}
\end{Schunk}

\item A maximum of 5 cohorts in total should be evaluated in the platform trial and for every patient we want the probability to start a new cohort to be 2\%. We want to trial to stop immediately after one successful combination therapy has been identified. The interim analysis should be conducted for every cohort after 50 patients and the final analysis after 100 patients.

\begin{Schunk}
\begin{Sinput}
R> cohorts_max <- 5 # maximum number of cohorts
R> cohort_random <- 0.02 # random cohort inclusion probability
R> # after how many successful cohorts should the platform trial stop?
R> sr_drugs_pos <- 1
R> n_int <- 50 # interim sample size
R> n_fin <- 100 # final sample size
\end{Sinput}
\end{Schunk}

\item In terms of response rates of the different arms, we want to specify the following discrete probability distributions:
   \begin{enumerate}
   \item  For the combination therapy: $P(\pi_{Comb}=0.35) = 0.4$, $P(\pi_{Comb}=0.40) = 0.4$, $P(\pi_{Comb}=0.45) = 0.2$.
    \item For monotherapy A: $P(\pi_{MonoA}=0.15) = 0.2$, $P(\pi_{MonoA}=0.20) = 0.4$, $P(\pi_{MonoA}=0.25) = 0.4$.
    \item For monotherapy B: $P(\pi_{MonoB}=0.15) = 0.3$, $P(\pi_{MonoB}=0.20) = 0.4$, $P(\pi_{MonoB}=0.25) = 0.3$.
    \item For SoC: $P(\pi_{SoC}=0.10) = 0.25$, $P(\pi_{SoC}=0.12) = 0.5$, $P(\pi_{SoC}=0.14) = 0.25$.
    \end{enumerate}

\begin{Schunk}
\begin{Sinput}
R> # Should response rates be drawn randomly?
R> random <- TRUE
R> 
R> # We specify the absolute response rates
R> random_type <- "absolute"
R> 
R> # What are the possible response rates for the
R> # combination therapies and with what probabilities
R> # should they be drawn?
R> rr_comb <- c(0.35, 0.40, 0.45)
R> prob_comb_rr <- c(0.4, 0.4, 0.2)
R> 
R> # What are the possible response rates for the
R> # monotherapy A and with what probabilities
R> # should they be drawn?
R> rr_mono <- c(0.15, 0.20, 0.25)
R> prob_mono_rr <- c(0.2, 0.4, 0.4)
R> 
R> # What are the possible response rates for the
R> # monotherapy B and with what probabilities
R> # should they be drawn?
R> rr_back <- c(0.15, 0.20, 0.25)
R> prob_back_rr <- c(0.3, 0.4, 0.3)
R> 
R> # What are the possible response rates for the
R> # SoCs and with what probabilities
R> # should they be drawn?
R> rr_plac <- c(0.10, 0.12, 0.14)
R> prob_plac_rr <- c(0.25, 0.5, 0.25)
\end{Sinput}
\end{Schunk}

Please note that vectors of arbitrary length can be specified here, which also allows approximation of continuous distributions.

\item At interim, we want to use the same endpoint as at the final analysis. We can achieve this in the following way:

\begin{Schunk}
\begin{Sinput}
R> # Set within-patient correlation options
R> # between final and interim outcome
R> rr_transform <- list(
+    function(x) {
+      return(c((1 - x), 0, 0, x))
+    }
+  )
R> # With which probabilities should these options be drawn?
R> prob_rr_transform <- 1
\end{Sinput}
\end{Schunk}

\end{itemize}

\subsection{Platform Trial Simulation}

After specifying all the required simulation parameters, we can simulate a single trial trajectory and save the results in an object \code{results\textunderscore ex1}.

\begin{Schunk}
\begin{Sinput}
R> set.seed(25)
R> 
R> results_ex1 <-
+    simulate_trial(
+      random = random, rr_comb = rr_comb, rr_mono = rr_mono,
+      rr_back = rr_back, rr_plac = rr_plac, rr_transform = rr_transform,
+      prob_rr_transform = prob_rr_transform, random_type = random_type,
+      prob_comb_rr = prob_comb_rr, prob_mono_rr = prob_mono_rr,
+      prob_back_rr = prob_back_rr, prob_plac_rr = prob_plac_rr,
+      safety_prob = safety_prob, cohort_random = cohort_random,
+      cohorts_max = cohorts_max, sr_drugs_pos = sr_drugs_pos,
+      target_rr = target_rr, sharing_type = sharing_type, n_int = n_int,
+      n_fin = n_fin, Bayes_Sup = Bayes_Sup, Bayes_Fut = Bayes_Fut
+  )
\end{Sinput}
\end{Schunk}

The return element by default contains two sublists, \code{results\textunderscore ex1\$Trial\textunderscore Overview} and \code{results\textunderscore ex1\$Stage\textunderscore Data}. In the first list element, we find overall summary information regarding the simulated platform design, such as the total number of patients:

\begin{Schunk}
\begin{Sinput}
R> results_ex1$Trial_Overview$Total_N
\end{Sinput}
\begin{Soutput}
[1] 272
\end{Soutput}
\end{Schunk}

The second list element contains detailed, stage-wise information on how the platform trial evolved over time, which responses were observed, which decisions were made, etc.

It is possible to plot the platform trial trajectory using the \code{plot\textunderscore trial()} function. The function is very simple: Assuming that the \code{simulate\textunderscore trial()} function has already been used to create a trial object, the \code{plot\textunderscore trial()} function can be applied directly to it. The result is a plotly interactive ggplot. The plot is arranged in a 3x2 grid: The plot on the top left gives an overview of the simulated study. The plot on the top right shows the simulated correlation of the final and interim outcomes, which for our particular simulation is 1. For better visibility, the points are jittered within the four quadrants. The plot in the middle on the left shows the empirical biomarker and histology response rates for the combination treatment, as well as the true histology response rate with respect to cohort. The plot in the middle on the right shows the empirical biomarker and histology response rates for the add-on monotherapy / monotherapy A treatment, as well as the true histology response rate with respect to cohort. The plot on the bottom left shows the empirical biomarker and histology response rates for the backbone monotherapy / monotherapy B, as well as the true histology response rate with respect to cohort. The plot on the bottom right shows the empirical biomarker and histology response rates for the combination treatment, as well as the true histology response rate with respect to cohort. Please note that - since calling this function produces an interactive figure created using \pkg{plotly} - the output cannot be directly shown in this PDF article, but rather we provide a screenshot of the created figure.

\begin{Schunk}
\begin{Sinput}
R> # plot_trial(results_ex1, unit = "n")
\end{Sinput}
\end{Schunk}

\begin{figure}[ht]
\begin{center}
\includegraphics[width=1\linewidth]{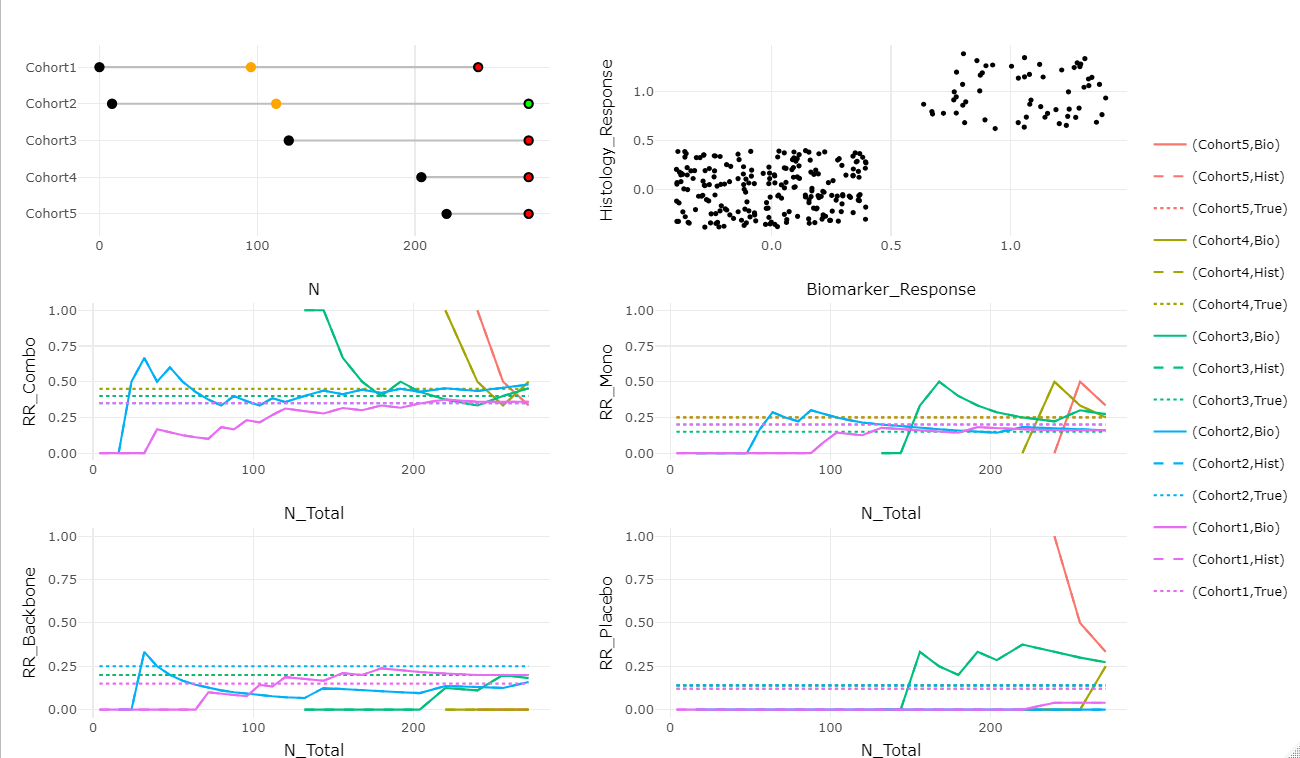}
\end{center}
\caption{\label{fig:ex1} Output of the \code{plot\textunderscore trial()} function for the Bayesian decision rules example.}
\end{figure}

\subsection{Computing Trial Operating Characteristics}

Usually, we are not interested in simulating one trajectory of a platform trial, but rather simulate many trajecories and compute the trial's operating characteristics. The \pkg{CohortPlat} packge has a dedicated function for this task. The \code{trial\textunderscore ocs()} function takes the same arguments as the \code{simulate\textunderscore trial()} function, which specify the desired study design, and additionally a few variables which define how many simulations will be run, on how many parallel cores the computation should be performed, whether to save the results as an Excel file or RData file and if so, where to save it. For the purposes of a short demonstration, we will use no parallel computing, run only very few iterations and not save the results anywhere else but in the \code{trial\textunderscore ocs\textunderscore ex1} object.

\begin{Schunk}
\begin{Sinput}
R> set.seed(25)
R> 
R> trial_ocs_ex1 <-
+    trial_ocs(
+      random = random, rr_comb = rr_comb, rr_mono = rr_mono,
+      rr_back = rr_back, rr_plac = rr_plac, rr_transform = rr_transform,
+      prob_rr_transform = prob_rr_transform, random_type = random_type,
+      prob_comb_rr = prob_comb_rr, prob_mono_rr = prob_mono_rr,
+      prob_back_rr = prob_back_rr, prob_plac_rr = prob_plac_rr,
+      safety_prob = safety_prob, cohort_random = cohort_random,
+      cohorts_max = cohorts_max, sr_drugs_pos = sr_drugs_pos,
+      target_rr = target_rr, sharing_type = sharing_type, n_int = n_int,
+      n_fin = n_fin, Bayes_Sup = Bayes_Sup, Bayes_Fut = Bayes_Fut,
+      # additional parameters compared to simulate_trial()
+      save = FALSE, # should results be saved locally as excel or Rdata
+      ret_list = TRUE, # should results be returned by the function
+      iter = 10 # number of iterations
+  )
\end{Sinput}
\end{Schunk}

The return object consists of two sublists: The first corresponds to an overview of the simulation parameters set and the second list element contains all computed operating characteristics, e.g. the disjunctive power, which is the probability to have at least one true positive decision in the platform trial (ref table \ref{tab:ocs}):

\begin{Schunk}
\begin{Sinput}
R> trial_ocs_ex1[[2]]$Disj_Power
\end{Sinput}
\begin{Soutput}
[1] 0.125
\end{Soutput}
\end{Schunk}

Please see table \ref{tab:ocs} in the appendix for an overview of all the computed operating characteristics.

\subsection{Running multiple simluation scenarios}

When conducting extensive simulation studies, usually the aim is to investigate the operating characteristics of a particular trial design across multiple different parameter options. As shown previously, the \pkg{CohortPlat} package facilitates calculation of operating characteristics for a fixed choice of simulation parameters. Using the \code{expand.grid()} function, we can easily create a data frame where each row corresponds to one set of parameter options. In order to achieve this, for every simulation parameter that is in fact a scalar, we specify a vector containing all values we want to investigate.

\begin{Schunk}
\begin{Sinput}
R> # Fixed simulation parameters
R> 
R> random                <- TRUE
R> random_type           <- "absolute"
R> rr_comb               <- 0.4
R> prob_comb_rr          <- 1
R> rr_mono               <- 0.2
R> prob_mono_rr          <- 1
R> rr_back               <- 0.2
R> prob_back_rr          <- 1
R> rr_plac               <- 0.1
R> prob_plac_rr          <- 1
R> cohorts_max           <- 3
R> trial_struc           <- "all_plac"
R> safety_prob           <- 0
R> sharing_type          <- "all"
R> cohort_offset         <- 0
R> sr_drugs_pos          <- Inf
R> sr_pats               <- Inf
R> n_fin                 <- 300
R> prob_rr_transform     <- 1
R> 
R> # Simulation Parameters of which we explore the impact
R> 
R> cohort_random         <- c(0.01, 0.02, 0.03)
R> n_int                 <- c(50, 100, 150, 200)
R> 
R> # Expand Grid
R> 
R> scenarios <- expand.grid(
+    random                = random,
+    random_type           = random_type,
+    rr_comb               = rr_comb,
+    prob_comb_rr          = prob_comb_rr,
+    rr_mono               = rr_mono,
+    prob_mono_rr          = prob_mono_rr,
+    rr_back               = rr_back,
+    prob_back_rr          = prob_back_rr,
+    rr_plac               = rr_plac,
+    prob_plac_rr          = prob_plac_rr,
+    prob_rr_transform     = prob_rr_transform,
+    trial_struc           = trial_struc,
+    cohorts_max           = cohorts_max,
+    safety_prob           = safety_prob,
+    sharing_type          = sharing_type,
+    sr_drugs_pos          = sr_drugs_pos,
+    sr_pats               = sr_pats,
+    cohort_random         = cohort_random,
+    cohort_offset         = cohort_offset,
+    n_int                 = n_int,
+    n_fin                 = n_fin
+  )
\end{Sinput}
\end{Schunk}

We can now loop over the different scenarios and save the relevant results in a common matrix. Simulation parameters which have a more complex structure are added manually:

\begin{Schunk}
\begin{Sinput}
R> # create empty matrix for results
R> sim_results <- matrix(ncol = 4, nrow = nrow(scenarios))
R> colnames(sim_results) <- c("cohort_random", "n_int", "PTP", "Disj_Power")
R> 
R> set.seed(20)
R> 
R> # loop over simulation scenarios
R> for (i in 1:nrow(scenarios)) {
+    # get list of arguments for trial_ocs function, including iterations
+    args_full <- c(
+      as.list(scenarios[i,]),
+      Bayes_Sup     = list(Bayes_Sup),
+      Bayes_Fut     = list(Bayes_Fut),
+      rr_transform  = list(list(function(x) c((1 - x), 0, 0, x))),
+      iter          = 4000,
+      coresnum      = 1,
+      # can be de/increased to use more/less available CPU cores
+      save          = FALSE,
+      ret_list      = TRUE
+    )
+  
+    # run simulations
+    ocs <- do.call(trial_ocs, args_full)[[2]]
+  
+    # save relevant information
+    sim_results[i,] <-
+        c(
+          args_full$cohort_random,
+          args_full$n_int,
+          ocs$PTP,
+          ocs$Disj_Power
+          )
+  }
\end{Sinput}
\end{Schunk}

\begin{figure}[t!]
\begin{center}
\begin{Schunk}

\includegraphics[width=\maxwidth]{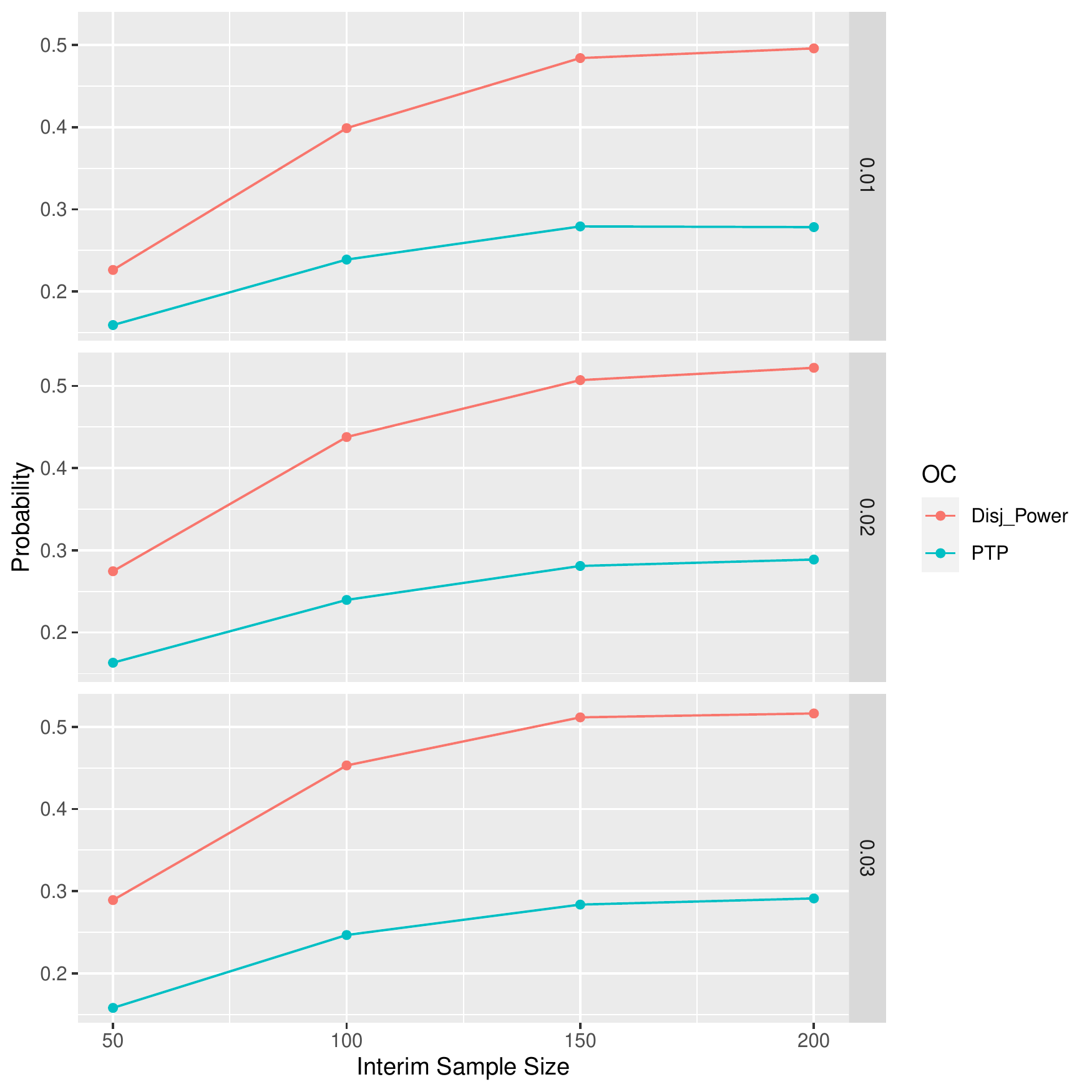} \end{Schunk}
\end{center}
\caption{\label{fig:ex3} Pair-wise and disjunctive power with respect to interim sample size (x-axis) and random cohort inclusion probability (rows).}
\end{figure}

Using \pkg{ggplot2}, we now visualize the impact of the chosen simulation parameters. Before that, we need to create a data frame suitable for usage. Results are shown in figure \ref{fig:ex3}.

\begin{Schunk}
\begin{Sinput}
R> results_df <-
+    as.data.frame(sim_results) %>%
+    pivot_longer(
+      cols = c("PTP", "Disj_Power"),
+      names_to = "OC",
+      values_to = "Probability"
+    )
R> 
R> ggplot(
+    data = results_df,
+    aes(x = n_int, y = Probability, col = OC)
+    ) +
+    geom_point() +
+    geom_line() +
+    facet_grid(rows = vars(cohort_random)) +
+    xlab("Interim Sample Size")
\end{Sinput}
\end{Schunk}

A similar, yet longer and more repetitive code can be used to also investigate the impact of simulation parameters that are vectors, matrices or lists.


\section{Summary and discussion} \label{sec:summary}

This article described the use of the \pkg{CohortPlat} package for simulation of cohort platform trials for combination therapies. To our knowledge, \pkg{CohortPlat} is the first package available on CRAN specifically aimed at simulating open-entry cohort platform trials. Difficulties of writing such software include the trade-off between modularity and feature richness in the context of highly flexible and adaptive clinical trials such as platform trials \citep{Meyer2021}.

The \pkg{CohortPlat} package focuses on including as many features of platform trials as possible, at the cost of modularity. This means that e.g. all current functions are aimed at binary endpoints - extending the existing software to include other types of endpoints would require substantial re-writing of the software code. Similarly, the cohort structure is fixed - if instead of two mono-therapies we wanted to include a third monotherapy, or include different dose-levels, again extensive re-writing of the software code would be necessary to the extent that the final product could be considered a separate package. We presented two different types of designs that can be simulated using the package (separate cohorts or cohorts with shared backbone monotherapy and SoC arms).

As for any simulation program, especially in a highly complex and dynamic context such as platform trials, simplifications were necessary. No patient level simulations were conducted. This leads to potentially slightly different sample sizes at interim and final than planned. However, this is also not uncommon in reality. Further simplifications included not simulating variations in recruitment speed due to availability of centers, external and internal events such as approval of competitor drugs or discontinuation of drug development programs. As the best possible compromise under uncertainty, we aimed to achieve a balanced randomization for every comparison in case of either no data sharing or sharing only concurrent data and refrained from investigating response-adaptive randomization, which might increase the efficiency of the platform trial \citep{Saville2016, Viele2020}.


\section*{Computational details}

The results in this paper were obtained using
\proglang{R}~4.1.2 with the
\pkg{CohortPlat}~1.0.3, \pkg{tidyr}~1.2.0, \pkg{dplyr}~1.0.7 and \pkg{ggplot2}~3.3.5 packages. The \pkg{CohortPlat} package uses several other packages within for e.g. parameter estimations, data management, interactive plotting and parallel computing, namely
    \pkg{dplyr},
    \pkg{purrr},
    \pkg{ggplot2},
    \pkg{plotly},
    \pkg{tidyr},
    \pkg{parallel},
    \pkg{doParallel},
    \pkg{foreach},
    \pkg{openxlsx},
    \pkg{forcats},
    \pkg{epitools} and \pkg{zoo}. \proglang{R} itself
and all packages used are available from the Comprehensive
\proglang{R} Archive Network (CRAN) at
\url{https://CRAN.R-project.org/}.

\section*{Acknowledgments}

EU-PEARL (EU Patient-cEntric clinicAl tRial pLatforms) project has received funding from the Innovative Medicines Initiative (IMI) 2 Joint Undertaking (JU) under grant agreement No 853966. This Joint Undertaking receives support from the European Union’s Horizon 2020 research and innovation program and EFPIA and Children’s Tumor Foundation, Global Alliance for TB Drug Development non-profit organization, Springworks Therapeutics Inc. This publication reflects the authors’ views. Neither IMI nor the European Union, EFPIA, or any Associated Partners are responsible for any use that may be made of the information contained herein. The research of Elias Laurin Meyer was funded until 11/2020 by Novartis through the University and not at an individual level.

\bibliography{refs}

\begin{appendix}

\section{Treatment Efficacy Assignment}

As discussed in section \ref{sec:assumptions_parameters}, treatment efficacy can be specified directly in responder rates or relative to SoC via risk-differences, risk-ratios and odds-ratios. In case of direct specification, for every cohort the responder rates of every treatment are drawn independently from the specified distributions. In case of specification relative to SoC, the following steps are performed independently for every cohort (in case a cohort does not include a SoC arm, the responder rates for the monotherpies and combination therapy are still derived in this way):

Firstly the responder rate for SoC, $\pi_{SoC}$, for a particular cohort is randomly drawn from the specified distribution $T_{SoC}$:

$$
\begin{array}{l}
\pi_{SoC} \sim T_{SoC}
 \end{array}
$$

In a second step, the relative effects of both of the monotherapies, $\gamma_{MonoA}$ and $\gamma_{MonoB}$, are randomly drawn from the specified distributions $T_{MonoA}$ and $T_{MonoB}$. The responder rates $\pi_{MonoA}$ and $\pi_{MonoB}$ are then derived as follows:

$$
\begin{array}{l}
\pi_{MonoA} = \pi_{SoC} \oplus \gamma_{MonoA}, \quad \gamma_{MonoA} \sim T_{MonoA} \\
\pi_{MonoB} = \pi_{SoC} \oplus \gamma_{MonoB}, \quad \gamma_{MonoB} \sim T_{MonoB}
 \end{array}
$$

where $\oplus$ is either an addition (in case of risk-differences), a multiplication (in case of risk-ratios) or a multiplication on the odds level (in case of odds-ratios). For the combination treatment, firstly the interaction term $\gamma_{Comb}$ is randomly drawn from the specified distribution $T_{Comb}$. The responder rate $\pi_{Comb}$ is then derived as follows:

$$
\begin{array}{l}
\pi_{Comb} = \pi_{SoC} \oplus \gamma_{MonoA} \oplus \gamma_{MonoB} \oplus \gamma_{Comb}, \quad \gamma_{Comb} \sim T_{Comb}
 \end{array}
$$

where $\gamma_{MonoA}$ and $\gamma_{MonoB}$ refer to the previously sampled treatment effects of monotherapies A and B and $\oplus$ is the same operation as before.

\section{Simulated Operating Characteristics}

\begin{longtable}{ p{0.2\linewidth} | p{0.75\linewidth}}
\caption{All operating characteristics computed by the \code{trial\_ocs()} function.} \\
\label{tab:ocs}
Operating characteristics & Definition \\ \hline
\hline
Avg Pat           & Average number of patients per platform trial\\ \hline
Avg Pat Comb      & Average number of patients on combination arms per platform trial\\ \hline
Avg Pat Mono      & Average number of patients on monotherapy A arms per platform trial\\ \hline
Avg Pat Back      & Average number of patients on monotherapy B arms per platform trial\\ \hline
Avg Pat Plac      & Average number of patients on SoC arms per platform trial\\ \hline
Avg RR Comb       & Average response rate of combination treatment arms across all platform trials\\ \hline
Avg RR Mono       & Average response rate of monotherapy A treatment arms across all platform trials\\ \hline
Avg RR Back       & Average response rate of monotherapy B arms across all platform trials\\ \hline
Avg RR Plac       & Average response rate of SoC treatment arms across all platform trials\\ \hline
SD RR Comb        & Standard deviation of response rate of combination treatment arms across all platform trials\\ \hline
SD RR Mono        & Standard deviation of response rate of monotherapy A treatment arms across all platform trials\\ \hline
SD RR Back        & Standard deviation of response rate of monotherapy B treatment arms across all platform trials\\ \hline
SD RR Plac        & Standard deviation of response rate of SoC treatment arms across all platform trials\\ \hline
Avg Suc Hist      & Average number of responders at final analysis per platform trial\\ \hline
Avg Suc Hist Comb & Average number of responders at final analysis on combination arms per platform trial\\ \hline
Avg Suc Hist Mono & Average number of responders at final analysis on monotherapy A arms per platform trial\\ \hline
Avg Suc Hist Back & Average number of responders at final analysis on monotherapy B arms per platform trial\\ \hline
Avg Suc Hist Plac & Average number of responders at final analysis on SoC arms per platform trial\\ \hline
Avg Suc Bio       & Average number of responders at interim analysis per platform trial\\ \hline
Avg Suc Bio Comb  & Average number of responders at interim analysis on combination arms per platform trial\\ \hline
Avg Suc Bio Mono  & Average number of responders at interim analysis on monotherapy A arms per platform trial\\ \hline
Avg Suc Bio Back  & Average number of responders at interim analysis on monotherapy B arms per platform trial\\ \hline
Avg Suc Bio Plac  & Average number of responders at interim analysis on SoC arms per platform trial\\ \hline
Avg Cohorts       & Average number of cohorts per platform trial\\ \hline
Avg TP            & Average number of true positives per platform trial, i.e. on average, for how many cohorts, which are in truth superior according to the defined target product profile, did the decision rules lead to a declaration of superiority\\ \hline
Avg FP            & Average number of false positives per platform trial, i.e. on average, for how many cohorts, which are in truth futile according to the defined target product profile, did the decision rules lead to a declaration of superiority\\ \hline
Avg TN            & Average number of true negatives per platform trial, i.e. on average, for how many cohorts, which are in truth futile according to the defined target product profile, did the decision rules lead to a declaration of futility\\ \hline
Avg FN            & Average number of false negatives per platform trial, i.e. on average, for how many cohorts, which are in truth superior according to the defined target product profile, did the decision rules lead to a declaration of futility\\ \hline
FDR               & "False Discovery Rate", the ratio of the sum of false positives (i.e. for how many cohorts, which are in truth futile according to the defined target product profile, did the decision rules lead to a declaration of superiority) among the sum of all positives (i.e. for how many cohorts did the decision rules lead to a declaration of superiority) across all platform trial simulations\\ \hline
PTP               & "Per-Treatment-Power", the ratio of the sum of true positives (i.e. for how many cohorts, which are in truth superior according to the defined target product profile, did the decision rules lead to a declaration of superiority) among the sum of all cohorts, which are in truth superior (i.e. the sum of true positives and false negatives) across all platform trial simulations, i.e. this is a measure of how wasteful the trial is with (in truth) superior therapies\\ \hline
PTT1ER            & "Per-Treatment-Type-1-Error", the ratio of the sum of false positives (i.e. for how many cohorts, which are in truth futile according to the defined target product profile, did the decision rules lead to a declaration of superiority) among the sum of all cohorts, which are in truth futile (i.e. the sum of false positives and true negatives) across all platform trial simulations, i.e. this is a measure of how sensitive the trial is in detecting futile therapies\\ \hline
FWER              & The proportion of platform trials, in which at least one false positive decision has been made, where only such platform trials are considered, which contain at least one cohort that is in truth futile\\ \hline
FWER BA           & The proportion of platform trials, in which at least one false positive decision has been made, regardless of whether or not any cohorts which are in truth futile exist in these platform trials\\ \hline
Disj Power        & The proportion of platform trials, in which at least one correct positive decision has been made, where only such platform trials are considered, which contain at least one cohort that is in truth superior\\ \hline
Disj Power BA     & The proportion of platform trials, in which at least one correct positive decision has been made, regardless of whether or not any cohorts which are in truth superior exist in these platform trials\\ \hline
Avg Cohorts     & Average number of cohorts per platform trial\\ \hline
Avg Perc Pat Sup Plac Th & Average percentage of patients on arms that are superior to SoC, whereby also arms that might have entered the platform are considered \\ \hline
Avg Perc Pat Sup Plac Real & Average percentage of patients on arms that are superior to SoC, taking into account only arms that were actually in the platform \\ \hline
Avg Pat Plac First Suc & Average number of patients on SoC until the first cohort was declared successful \\ \hline
Avg Pat Plac Pool & Average number of patients that could have been randomised to SoC \\ \hline
Avg Cohorts First Suc & Average number of cohorts until the first cohort was declared successful \\ \hline
Avg any P & Percentage of platform trials where any alternative hypothesis was true \\ \hline
Coh Safety STOP Perc & Percentage of cohorts that stopped for safety \\ \hline
Coh Int STOP Perc & Percentage of cohorts that stopped at interim for early futility \\ \hline
Coh Int GO Perc & Percentage of cohorts that stopped at interim for early efficacy \\ \hline
Avg Safety STOP Trial & Average number of cohorts that stopped for safety \\ \hline
Avg Int STOP Trial & Average number of cohorts that stopped at interim for early futility \\ \hline
Avg Int GO Trial & Average number of cohorts that stopped at interim for early efficacy \\ \hline
Avg PTP Trial & Similar to "PTP", however calculated on an individual platform trial level and then averaged \\ \hline
Avg PTT1ER Trial & Similar to "PTT1ER", however calculated on an individual platform trial level and then averaged \\ \hline
Avg FDR Trial & Similar to "FDR", however calculated on an individual platform trial level and then averaged \\ \hline
Dist FWER & Vector of FWER for all simulated platform trials \\ \hline
Dist FDR & Vector of FDR for all simulated platform trials \\ \hline
Dist PTT1ER & Vector of PTT1ER for all simulated platform trials \\ \hline
Dist PTP & Vector of PTP for all simulated platform trials \\ \hline
Dist Disj Power & Vector of Disj Power for all simulated platform trials \\ \hline
\hline
\end{longtable}

\section{Frequentist Decision Rules} \label{freq_dec}

Imagine we want to specify the following frequentist decision rules.

At interim, declare early futility for the combination therapy if any of the following conditions holds:
\begin{enumerate}
    \item The one-sided p-value from a Chi-Square Test without continuity correction comparing the combination therapy response rate against the monotherapy A response rate is above 0.5.
    \item The one-sided p-value from a Chi-Square Test without continuity correction comparing the combination therapy response rate against the monotherapy B response rate is above 0.5.
    \item The one-sided p-value from a Chi-Square Test without continuity correction comparing the monotherapy A response rate against the SoC response rate is above 0.5.
    \item The one-sided p-value from a Chi-Square Test without continuity correction comparing the monotherapy B response rate against the SoC response rate is above 0.5.
\end{enumerate}

At final, declare superiority for the combination therapy if all of the following conditions hold:
\begin{enumerate}
    \item The one-sided, Bonferroni-corrected p-value from a Chi-Square Test without continuity correction comparing the combination therapy response rate against the monotherapy A response rate is below 0.10.
    \item The one-sided, Bonferroni-corrected p-value from a Chi-Square Test without continuity correction comparing the combination therapy response rate against the monotherapy B response rate is below 0.10.
    \item The one-sided, Bonferroni-corrected p-value from a Chi-Square Test without continuity correction comparing the monotherapy A response rate against the SoC response rate is below 0.10.
    \item The one-sided, Bonferroni-corrected p-value from a Chi-Square Test without continuity correction comparing the monotherapy B response rate against the SoC response rate is below 0.10.
\end{enumerate}

These frequentist decision rules are achieved by specifying the decision rules in the following way:

\begin{Schunk}
\begin{Sinput}
R> # Set decision rules ----------------
R> 
R> # Frequentist Futility Rules (only at interim)
R> 
R> # Interim
R> # Comparison Combination vs Monotherapy A
R> P_Fut1_Int <- list(list(testfun = function(x)
+    stats::prop.test(x, alternative = "less", correct = FALSE),
+    p_fut = 0.5, p_adj = "none"))
R> # Comparison Combination vs Monotherapy B
R> P_Fut2_Int <- list(list(testfun = function(x)
+    stats::prop.test(x, alternative = "less", correct = FALSE),
+    p_fut = 0.5, p_adj = "none"))
R> # Comparison Monotherapy A vs SoC
R> P_Fut3_Int <- list(list(testfun = function(x)
+    stats::prop.test(x, alternative = "less", correct = FALSE),
+    p_fut = 0.5, p_adj = "none"))
R> # Comparison Monotherapy B vs SoC
R> P_Fut4_Int <- list(list(testfun = function(x)
+    stats::prop.test(x, alternative = "less", correct = FALSE),
+    p_fut = 0.5, p_adj = "none"))
R> 
R> # Final Analysis (Decision Rules empty, but still need to be specified)
R> # Comparison Combination vs Monotherapy A
R> P_Fut1_Fin <- list(list(testfun = NA, p_fut = NA, p_adj = NA))
R> # Comparison Combination vs Monotherapy B
R> P_Fut2_Fin <- list(list(testfun = NA, p_fut = NA, p_adj = NA))
R> # Comparison Monotherapy A vs SoC
R> P_Fut3_Fin <- list(list(testfun = NA, p_fut = NA, p_adj = NA))
R> # Comparison Monotherapy B vs SoC
R> P_Fut4_Fin <- list(list(testfun = NA, p_fut = NA, p_adj = NA))
R> 
R> # Wrapup in package format
R> P_Fut <- list(list(P_Fut1_Int, P_Fut2_Int, P_Fut3_Int, P_Fut4_Int),
+                list(P_Fut1_Fin, P_Fut2_Fin, P_Fut3_Fin, P_Fut4_Fin))
R> 
R> # Frequentist Superiority Rules (only at final)
R> 
R> # Interim (Decision Rules empty)
R> # Comparison Combination vs Monotherapy A
R> P_Sup1_Int <- list(list(testfun = NA, p_sup = NA, p_prom = NA))
R> # Comparison Combination vs Monotherapy B
R> P_Sup2_Int <- list(list(testfun = NA, p_sup = NA, p_prom = NA))
R> # Comparison Monotherapy A vs SoC
R> P_Sup3_Int <- list(list(testfun = NA, p_sup = NA, p_prom = NA))
R> # Comparison Monotherapy B vs SoC
R> P_Sup4_Int <- list(list(testfun = NA, p_sup = NA, p_prom = NA))
R> 
R> # Final Analysis
R> # Comparison Combination vs Monotherapy A
R> P_Sup1_Fin <- list(list(
+  testfun = function(x) stats::prop.test(x, alternative = "less",
+                                         correct = FALSE),
+  p_sup = 0.10, p_prom = 0, p_adj = "B"))
R> # Comparison Combination vs Monotherapy B
R> P_Sup2_Fin <- list(list(
+  testfun = function(x) stats::prop.test(x, alternative = "less",
+                                         correct = FALSE),
+  p_sup = 0.10, p_prom = 0, p_adj = "B"))
R> # Comparison Monotherapy A vs SoC
R> P_Sup3_Fin <- list(list(
+  testfun = function(x) stats::prop.test(x, alternative = "less",
+                                         correct = FALSE),
+  p_sup = 0.10, p_prom = 0, p_adj = "B"))
R> # Comparison Monotherapy B vs SoC
R> P_Sup4_Fin <- list(list(
+  testfun = function(x) stats::prop.test(x, alternative = "less",
+                                         correct = FALSE),
+  p_sup = 0.10, p_prom = 0, p_adj = "B"))
R> 
R> # Wrapup in package format
R> P_Sup <- list(list(P_Sup1_Int, P_Sup2_Int, P_Sup3_Int, P_Sup4_Int),
+                list(P_Sup1_Fin, P_Sup2_Fin, P_Sup3_Fin, P_Sup4_Fin))
\end{Sinput}
\end{Schunk}

\end{appendix}


\end{document}